\documentclass{emulateapj}
\slugcomment{DRAFT  v.\today}

\shorttitle{}
\shortauthors{Compi\`egne et al.}

\begin{document}

\title{Dust in the diffuse emission of the galactic plane \\  
The {\it Herschel}/{\it Spitzer} SED fitting\footnote{{\it Herschel} 
is an ESA space observatory with science instruments provided by 
European-led Principal Investigator consortia and with important 
participation from NASA. Spitzer Space Telescope  is operated by the 
Jet Propulsion Laboratory, California Institute of Technology under 
a contract with NASA}}

   \author{M. Compi\`egne\altaffilmark{1},
                N. Flagey\altaffilmark{2},
                A. Noriega-Crespo\altaffilmark{2},
                P.~G. Martin\altaffilmark{1, 3},
                J.-P. Bernard\altaffilmark{4},                
                R. Paladini\altaffilmark{2},
                S. Molinari\altaffilmark{5}
        }
      \email{compiegne@cita.utoronto.ca}
      \altaffiltext{1}{Canadian Institute for Theoretical Astrophysics, 
University of Toronto, 60 St. George Street, Toronto, ON M5S 3H8, Canada}
      \altaffiltext{2}{Spitzer Science Center, California Institute of 
Technology, 1200 East California Boulevard, MC 220-6, Pasadena, CA 91125, USA}
      \altaffiltext{3}{Department of Astronomy \& Astrophysics, University 
of Toronto, 50 St. George Street, Toronto, ON M5S 3H4, Canada }
       \altaffiltext{4}{Centre d'Etude Spatiale des Rayonnements, 
CNRS et Universit\'e Paul Sabatier-Toulouse 3, Observatoire Midi-Pyr\'en\'ees,
 9 Av. du Colonel Roche, 31028 Toulouse Cedex 04, France}
      \altaffiltext{5}{INAF-Istituto Fisica Spazio Interplanetario, 
Via Fosso del Cavaliere 100, I-00133 Roma, Italy}

\begin{abstract}
  The first {\it Herschel} Hi-Gal images of the galactic plane unveil the 
far-infrared diffuse emission of the interstellar medium with an
unprecedented
angular resolution and sensitivity. In this paper, we present the first 
analysis of these data in combination with that of {\it Spitzer} 
GLIMPSE \& MIPSGAL. We selected a relatively diffuse and low excitation region
of the $l\sim59\degr$ Hi-Gal Science Demonstration Phase field to perform a 
pixel by pixel fitting of the 8 to 500\,$\mu$m SED using the {\tt DustEM}
dust emission model.
We derived maps of the Very Small Grains (VSG) and Polycyclic Aromatic 
Hydrocarbons (PAH) abundances from the model. Our analysis allows us to 
illustrate that the Aromatic Infrared Bands (AIB) intensity does not 
trace necessarily the PAH abundance but rather the product of 
``abundance $\times$ column density $\times$ intensity of the exciting 
radiation field''. We show that the spatial structure of PACS\,$70\,\mu$m map 
resembles the shorter wavelengths (e.~g. IRAC\,$8\,\mu$m) maps, because
they trace both the intensity of exciting radiation field and column 
density. We also show that the modeled VSG contribution to
PACS\,70\,$\mu$m (PACS\,160\,$\mu$m) band intensity can be up to 50\% (7\%). 
The interpretation of diffuse emission spectra at these wavelengths must 
take stochastically heated particles into account.
Finally, this preliminary study emphasizes the potential of analyzing the 
full dust SED sampled by {\it Herschel} and {\it Spitzer} data, with 
a physical dust model ({\tt DustEM}) to reach the properties of the dust at 
simultaneously large and small scales.
\end{abstract}



\section{Introduction}\label{sect:intro}

\begin{figure*}
    \centering
      \includegraphics[width=.97\textwidth, angle=0]{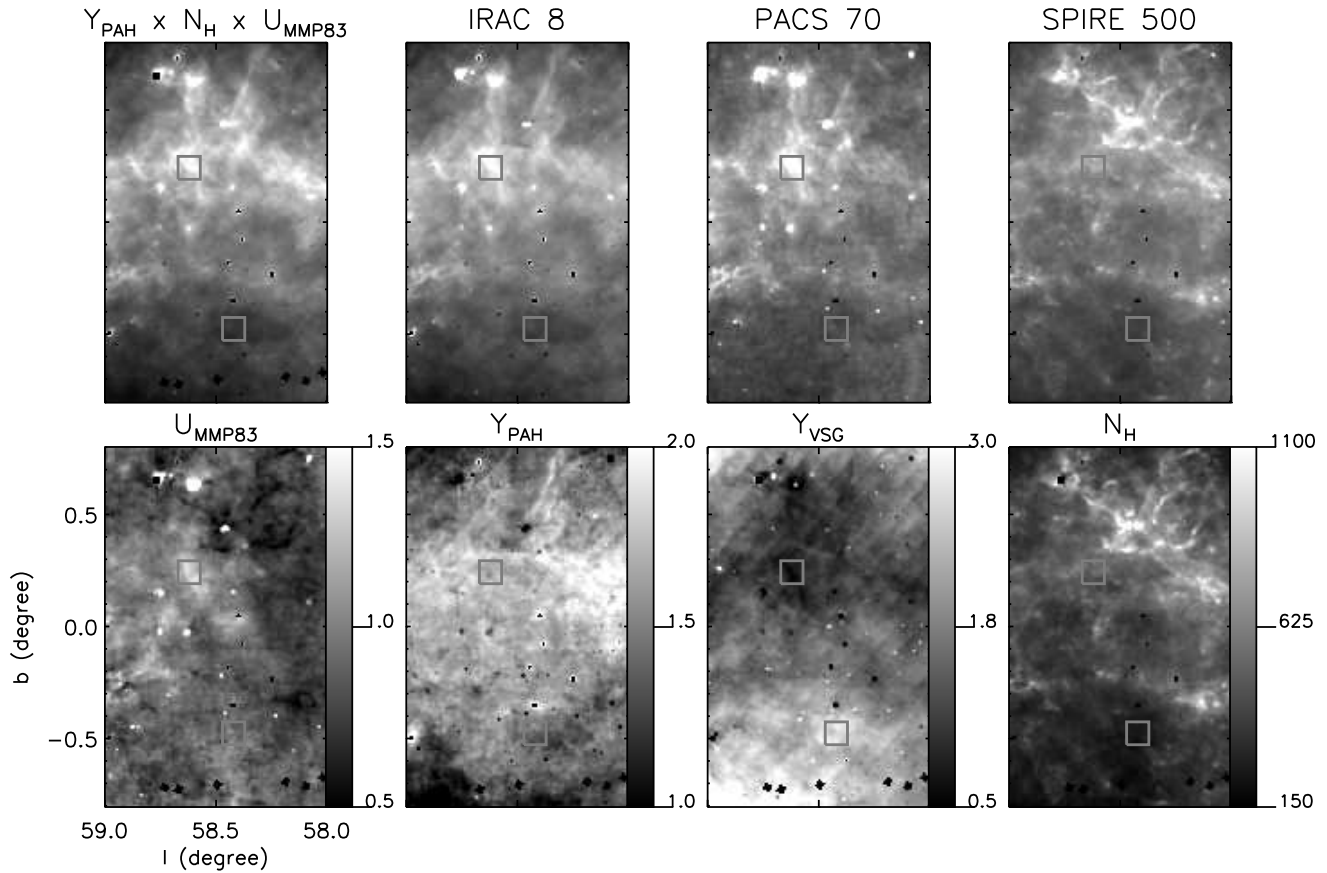}
      \caption{IRAC\,$8\,\mu$m, PACS\,$70\,\mu$m and SPIRE\,$\,500\mu$m maps.
              The obtained maps for the fitted parameters are $Y_{PAH}$, 
              $Y_{VSG}$, $N_H$ and $U_{MMP83}$. Also shown is a map of the 
              product  $Y_{PAH} \times N_H \times U_{MMP83}$.The two boxes 
              delineate the areas used to obtain the spectra of 
              Fig.\,\ref{fig:spectra}. Note that the black spots are due 
              to pixel masking (e.g. 8$\,\mu$m point sources).}
          \label{fig:fitres}
\end{figure*}
 
\begin{figure}
    \centering
      \includegraphics[width=.47\textwidth]{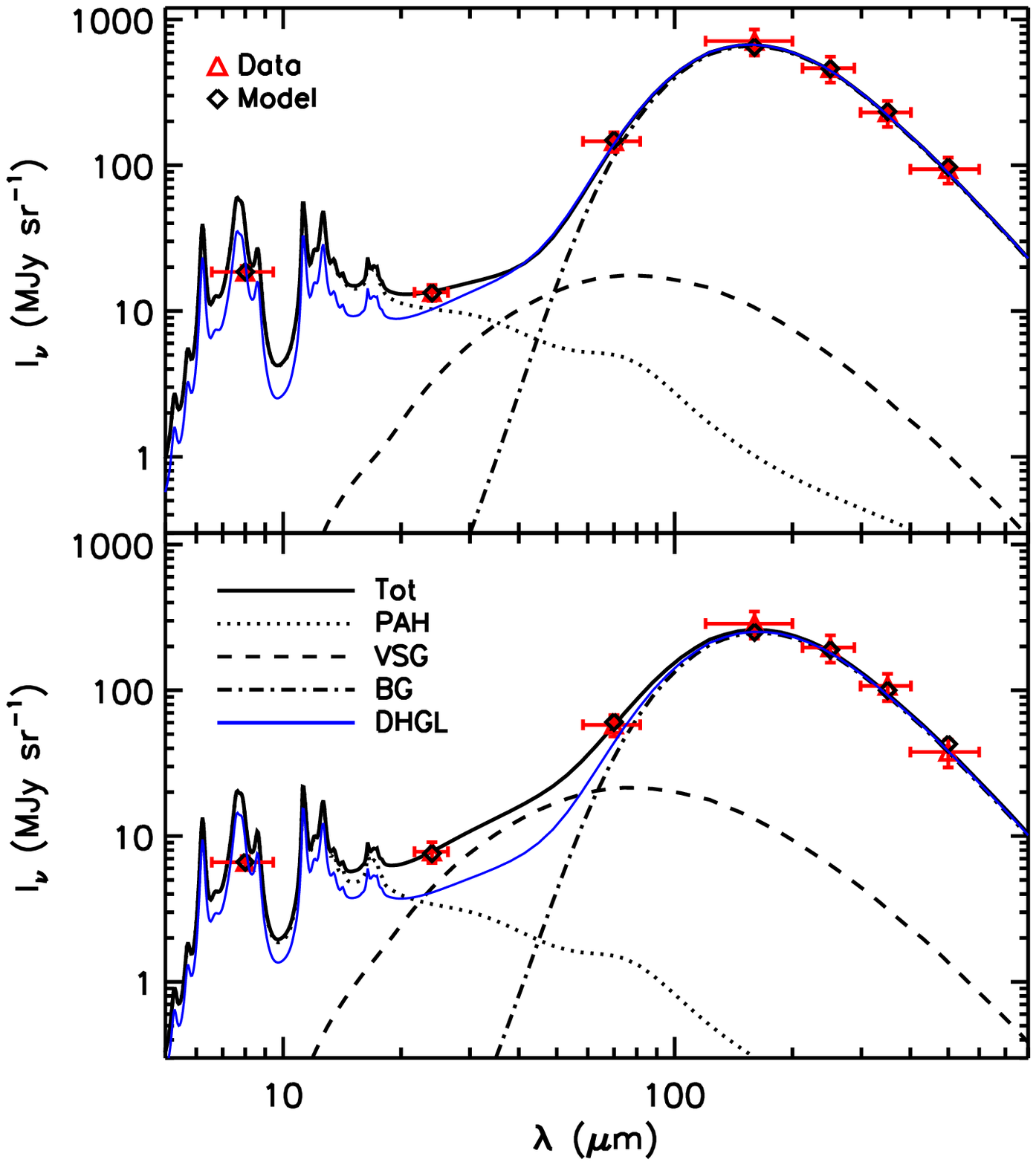}
      \caption{Mean spectra over the boxes shown on 
        Fig.\,\ref{fig:fitres} and corresponding fitted model.
        The upper (lower) spectrum correspond to the northern (southern) box.
        For comparison, the blue lines show the spectra obtain for the fitted 
        $N_H$ and $U_{MMP83}$ but
        for the reference diffuse high galactic latitude dust properties 
        (DHGL,  $Y_{PAH}$ and $Y_{VSG}\,=\,1$).
                    }
          \label{fig:spectra}
\end{figure}

The first {\it Herschel} \citep{pilbratt2010} images of the galactic plane 
obtained during the Science Demonstration Phase (SDP) reveal with an 
unprecedented beauty and detail the far-infrared diffuse emission of the 
interstellar medium (ISM).  The Hi-Gal project \citep{molinari2010} provides
an unbiased photometric survey of the inner Galactic plane emission between 
70 and 500\,$\mu$m where the spectral energy distribution (SED) is dominated 
by the largest dust grains ($a\sim0.1\,\mu$m).  Combined with the {\it Spitzer}
GLIMPSE \citep{churchwell2009} and MIPSGAL \citep{carey2009} surveys that 
cover the dust SED from 3.6 to 24\,$\mu$m, light mostly emitted by the 
smallest grains (a\,$\la 10\,$nm), we will for the first time reliably sample 
the full dust SED of the diffuse emission at all spatial scales down to 
$\sim40\arcsec$ and over a large fraction of the galactic plane.

At 70 and 160\,$\mu$m many of the observed structures show a close spatial 
correlation with some of the classical infrared tracers of small dust 
particles, like the emission measured at 8 and 24\,$\mu$m, and therefore, one 
expects a partially similar origin and/or a similar dependence to ISM 
physical properties. To properly understand the origin of this emission is 
the first step of the data analysis.

Dust is heated by stellar ultraviolet(UV)-visible photons and re-radiate the 
absorbed energy in the infrared.  Assuming some dust properties, this dust 
emission is widely used to trace crucial quantities like cloud masses or star 
forming activity. However, to derive accurately these quantities requires to 
take into account the dust properties evolution, how it affects its emission 
and how it reflects the ISM properties. On the other hand, dust plays a 
crucial role for the ISM physics and chemistry. 
Since dust and the rest of ISM are tightly interlock, changes on the dust properties
affect those of the gas phase.
It is then necessary to characterize the physical processes responsible for the
dust evolution to assess their role within the ISM lifecycle and to be
able to use dust as a reliable tracer of the ISM. Accessing the full dust 
SED over a large fraction of the sky allows to follow the behavior of every 
dust component over a broad range of physical conditions which is an invaluable
information for the study of the dust evolution.

In this paper, we present the first analysis of the Hi-Gal data combined 
with that of GLIMPSE and MIPSGAL to obtain the full dust SED from 
8 to 500\,$\mu$m. In the section\,\ref{sect:obs}, we describe the 
observations and the fitting method that makes use of a physical dust model, 
{\tt DustEM}\footnote{{\tiny To be found at {\tt
      http://www.ias.u-psud.fr/DUSTEM}}} 
\citep{compiegne2010}. In the section\,\ref{sect:results}, we 
present and discuss the results and how our analysis method allows us to 
better interpret the behavior of extended emission observed with the 
{\it Herschel/Spitzer} photometers. We conclude in the section\,\ref{sect:conclusion}.

\section{Observations \& analysis method}\label{sect:obs}

Two fields were observed (at $l\sim30\degr$ and $l\sim59\degr$) during the SDP
 as part of the Hi-Gal program with PACS\,70 and 160\,$\mu$m and all the SPIRE
 channels \citep{molinari2010a}.  For this study, to satisfy the assumptions 
made in the model we describe below, we focus on a subfield of the 
$l\sim59\degr$ field that shows no obvious HII regions or young stars cluster 
and that is relatively diffuse regarding the rest of the $l\sim59\degr$ field 
and the $l\sim30\degr$ field.  We also avoid the higher latitudes 
($|b|\ga0.8\degr$) that display weak intensity where {\it Herschel} 
data could be less reliable at this early stage of data processing.  
The maps were obtained with the ROMAGAL pipeline (Traficante et al., 2010, 
MNRAS, submitted).  
For PACS\,160\,$\mu$m, SPIRE\,250, 350 and 500\,$\mu$m, we applied the 
calibration described in \citet{bernard2010}. The gain uncertainty for 
these data is taken to be 20\%. The zero level was corrected by 
cross-calibration with Planck data with absolute uncertainties 
of $\pm$\,19.8, 14.6, 7.5 and 3.0\,$\rm{MJy\,sr^{-1}}$ at 160, 250, 350 and 
500\,$\mu$m, respectively, which represent about half of the gain uncertainty 
for the faintest pixels of the studied region.  
We cross-calibrate the PACS\,70\,$\mu$m data with the MIPS\,70\,$\mu$m data 
from the MIPSGAL survey (Paladini et al., 2010, in prep).  We then apply the 
15\% gain uncertainty of MIPS\,70\,$\mu$m.  We also use the 
IRAC\,$8\,\mu$m data from the GLIMPSE survey and the MIPS\,$24\,\mu$m data 
from the MIPSGAL survey. The gain uncertainty for these two dataset is 
taken to be 10\%.  Point sources are subtracted from the IRAC\,$8\,\mu$m data.

The zodiacal light contribution, important at shorter infrared wavelengths 
\citep{kelsall98} has been removed from all the {\it Spitzer} data 
(on average, 1.1, 16.5 and 4.5\,MJy\,sr$^{-1}$ at 8, 24 and 70\,$\mu$m,
respectively) and then from the PACS\,70\,$\mu$m following its 
cross-calibration on MIPS\,70\,$\mu$m.  We do not perform any subtraction of 
the zodiacal emission at longer wavelenghts where its contribution is 
negligible ($<$\,1\,MJy\,sr$^{-1}$). We bring every map to the lowest 
resolution of the SPIRE\,500\,$\mu$m one (FWHM$\,\sim37\arcsec$). We do 
this assuming a Gaussian Point Spread Function of appropriate width.  
We then project all maps into the SPIRE\,500\,$\mu$m grid (pixel field 
of view$\,\sim11.5\arcsec$).  Fig.\,\ref{fig:fitres} displays the IRAC\,
$8\,\mu$m, PACS\,70\,$\mu$m and SPIRE\,500\,$\mu$m maps.

We use the {\tt DustEM} dust model described in \citet{compiegne2010} to 
analyse the data. We consolidated the standard four grain populations 
into three: the Polycyclic Aromatic Hydrocarbons (PAH), 
Small Amorphous Carbons representing the Very Small Grains (VSG) and we 
merged Large Amorphous Carbons and Silicates into a single Big Grains (BG) 
population.  
Using the {\sc mpfit} \citep{markwardt2009} IDL minimization 
routine\footnote{{\tiny To be found at {\tt http://purl.com/net/mpfit}}}, 
we choose to adjust the following four parameters to fit the observed SED for 
each pixel: the (i) PAH and (ii) VSG abundances relative to BG, $Y_{PAH}$ 
and $Y_{VSG}$, (iii) the BG opacity, $\tau_{BG}$, and (iv)\,$U_{MMP83}$ a 
scaling factor of the ``solar neighborhood'' \citet{mathis83} (hereafter 
MMP83) exciting radiation field.
Self-extinction along the line of sight can be significant at 8\,$\mu$m and 
is accounted for assuming 
$I_\lambda\,=\,I_{0,\lambda}\,\frac{1-e^{-\tau_\lambda}}{\tau_\lambda}$ 
where $I_{0,\lambda}$ is the integrated emissivity and $\tau_\lambda$ 
the total dust opacity that is computed from the dust model using $Y_{PAH}$, 
$Y_{VSG}$ and $\tau_{BG}$. 
Since in this first analysis we do not perform 
any separation of the components along the line of sight,
the derived parameters result from the spatial mixing of different
physical conditions. Fortunately, at $l\sim59\degr$, we expect only 
contributions from the Vulpecula star formation region (d$\sim$2\,kpc) and 
the Perseus arm (d$\sim$8.5\,kpc).

We focus on the smallest particles behavior (PAH and VSG) and hence, we do 
not assume any variation of the BG properties (i.e. $\tau_{BG}/N_H$\,=\, 
constant).  In that case and to ease the following discussion, we can 
convert $\tau_{BG}$ into the hydrogen column density, $N_H$, and assume 
the relative abundances of PAH and VSG, $Y_{PAH}$ and $Y_{VSG}$ to be their 
abundance relative to hydrogen.  Indeed, emissivity and/or abundance of the 
BG is known to evolve \citep[e.g.][]{stepnik2003, desert2008} and can be 
studied using the {\it Herschel} data \citep[see][]{paradis2010} but to 
take such variations into account is beyond the scope of this paper.

Fig.\,\ref{fig:spectra} illustrates the obtained fitted spectra for two 
observed SEDs.  The photometric points computed from a modeled spectrum take 
into account the color corrections.  Schematically, the 160, 250, 350 
and 500\,$\mu$m photometric points give constraints on $U_{MMP83}$ and 
$\tau_{BG}$ ($N_H$) through the BG emission. The shape of the BG SED depends 
on $U_{MMP83}$ as this grain population is at thermal equilibrium, in contrast
with the stochastically heated grains (PAH and VSG) whose SED shape is 
invariant regarding the radiation field intensity \citep[see][]{draine2007, 
compiegne2010}. Hence, given $U_{MMP83}$ the absolute level of the BG SED 
constrains $N_H$.  Finally, the 8 and 24\,$\mu$m constrain the abundances of 
the two stochastically heated population since the intensity of their emission
scales as $Y \times N_H \times U_{MMP83}$.

\section{Results \&  discussion}\label{sect:results}

Fig.\,\ref{fig:fitres} shows maps of the $Y_{PAH}$, $Y_{VSG}$, $N_H$ 
and $U_{MMP83}$ parameters and a map of the product 
$Y_{PAH} \times N_H \times U_{MMP83}$. $\chi^2_{reduced}\la2$ for all pixels 
over these maps. $N_H$ is given in unit of $10^{20}\,H\,cm^{-2}$ and 
$U_{MMP83}$ is dimensionless. $Y_{PAH}$ and $Y_{VSG}$ are given relative 
to the value for the diffuse high galactic latitude medium 
(DHGL, $|b| \ga 15\degr$), $M_{PAH}/M_{H}\,=\,7.8\,10^{-4}$ and 
$M_{VSG}/M_{H}\,=\,1.65\,10^{-4}$ \citep[see][]{compiegne2010}. 
Notice that $M_{BG}/M_{H}\,=\,9.25\,10^{-3}$.
The uncertainties on these parameters are given as computed in {\sc mpfit} 
from the covariance matrix.

The column density lies between $\sim$$1.5\,10^{22}\,H\,cm^{-2}$ for the most 
diffuse part and $\sim$$1.1\,10^{23}\,H\,cm^{-2}$ toward the dense filamentary
structures, with a relative uncertainty of $\sim$18\%.
The $U_{MMP83}$ ranges between $\sim$0.5 and $\sim$1.5 with a relative 
uncertainty of $\sim$22\%, which is consistent with the absence of young 
stellar cluster over our field. In agreement with the fact that the radiation 
field is shielded, we see that $U_{MMP83} $ decreases toward the highest 
column density regions. These $U_{MMP83}$ and $N_H$ values are in agreement 
with previous estimates \citep[see][]{bernard2010}.  
$Y_{PAH}$ goes from $\sim$1.0 to $\sim$2.0 with a relative uncertainty of 
$\sim$16\%. The obtained $Y_{VSG}$ spans a wider range than $Y_{PAH}$, 
going from $\sim$0.5 to $\sim$3.0 but with a relative uncertainty going from 
$\sim$40\% for the highest values to $\sim$60\% for the smallest values.
The variations of these parameters over the map are therefore 
significant.

Fig.\,\ref{fig:spectra} shows the averaged SED over the two boxes seen on 
Fig.\,\ref{fig:fitres} and illustrate the wide $Y_{VSG}$ variations.
For the top (bottom) panel spectrum we have 
$N_{H}\,=\, 5.1$$\pm$$0.9\,10^{22} \,H\,cm^{-2}$ 
($2.5$$\pm$$0.5\,10^{22}\,H\,cm^{-2}$), 
$U_{MMP83}\,=\, 1.2$$\pm$$0.2$ ($0.9$$\pm$$0.2$),
$Y_{PAH}\,=\,1.7$$\pm$$0.2$ ($1.4$$\pm$$0.2$)
and $Y_{VSG}\,=\,0.8$$\pm$$0.5$  ($2.7$$\pm$$1.0$). 

$Y_{PAH}$/$Y_{VSG}$ seems to  vary at large spatial scales (decreases from 
$b\sim0.2\degr$ to $b\sim-0.7\degr$)
and also at the edge of some dense filaments (near the ``chimney'' region 
around coordinate 58.4, +0.55). 
Both $Y_{PAH}$ and $Y_{VSG}$ decrease towards some of the densest filamentary
structures but it appears not to be systematic and could be biased
by the too simplistic assumption made on the self-absorption.
A decrease of the smallest dust abundance towards dense regions regarding 
the biggest grains was reported by previous work and was interpreted as the 
coagulation of these smallest particles together with the bigger ones 
\citep[e.g.][]{stepnik2003, flagey2009}.  
Previous studies also reported a lack of correlation or even an 
anti-correlation between the aromatic infrared bands and the mid-IR continuum 
emission (interpreted as the evolution of small dust properties) at the 
illuminated ridge of molecular clouds \citep[e.g.][]{abergel2002, berne2007, 
compiegne2008}, toward high galactic cirrus at the interface between atomic 
and molecular material \citep[][]{mamd2002} or at galactic scale in the Large 
Magellanic Cloud \citep{paradis2009}.

The spatial structure of SPIRE\,$500\,\mu$m intensity, that is dominated
by the BG contribution, is well correlated with $N_H$.  Indeed, $U_{MMP83}$ 
and subsequently the BG temperature is quite stable in our field so that the 
intensity variations at these wavelengths (Rayleigh tail of the blackbody 
like emission) are dominated by the column density variations (assuming no 
BG emissivity variations).
Comparing the $8\,\mu$m and $Y_{PAH}$ maps, it is striking that the AIB 
intensity does not directly trace the PAH abundance. PAH are stochastically 
heated and the observed AIB intensity (i.e. IRAC\,$8\,\mu$m) then scales 
linearly with the product $Y_{PAH} \times N_H \times U_{MMP83}$ as seen on 
Fig.\,\ref{fig:fitres}.  The only difference is related to the extinction 
along the line of sight that can be seen in the IRAC\,$8\,\mu$m (as dark 
filamentary structures correlated with the $N_H$ map) and not in the 
$Y_{PAH} \times N_H \times U_{MMP83}$ map.
The PACS\,70\,$\mu$m image shows a close spatial correlation with the 
classical infrared tracers of small dust particles, like the emission 
measured at 8\,$\mu$m (see Fig.\,\ref{fig:fitres}).  The emission at 
70\,$\mu$m is not due to a single dust component.  As seen on 
Fig\,\ref{fig:spectra}, in the framework of our model both the VSG that 
are stochastically heated and the BG that are at thermal equilibrium 
contribute to the emission at this wavelength.  For $U_{MMP83}\la100$, the 
BG emission at $70\,\mu$m fall in the Wien part of the blackbody like emission
 that makes it more sensitive to $U_{MMP83}$ than at 500\,$\,\mu$m.  
Therefore, the two emission components (from VSG and BG) at $70\,\mu$m are 
sensitive to both $N_H$ and $U_{MMP83}$ explaining the better correlation of 
this map with the $8\,\mu$m map (also sensitive to both $N_H$ and $U_{MMP83}$)
than with the 500\,$\,\mu$m map (more sensitive to $N_H$).

The model allows us to compute the relative contribution of the three dust 
populations in the different photometric bands. 
The VSG contribution to PACS\,$70\,\mu$m increases if the VSG abundance 
relative to BG increases and/or if $U_{MMP83}$ decreases (shifting the BG 
emission toward longer wavelengths). In the studied field, the VSG 
contribution to the PACS\,$70\,\mu$m intensity goes from $\sim10\%$ up to 
$\sim50\%$ with a median value of $\sim27\%$. The maximum contribution of VSG 
to PACS$\,100\,\mu$m and PACS$\,160\,\mu$m is $\sim 17\%$ and $\sim 7\%$ 
(the median is $9\%$ and $3\%$), respectively. 
For the top (bottom) panel spectrum of Fig.\,\ref{fig:spectra}, the contribution is 12, 4, 2\%
(35, 13, 5\%) for PACS\,70, 100 and 160\,$\mu$m, respectively.
This result strongly 
suggests that the proper analysis of {\it Herschel} spectrum of diffuse 
emission including $70\,\mu$m may require to account for 
stochastically heated grains.

\section{Summary \& Conclusion}\label{sect:conclusion}

We have presented the first analysis of the diffuse emission of the galactic 
plane as observed by {\it Herschel} combining the Hi-Gal data with the 
GLIMPSE/MIPSGAL {\it Spitzer} data.
Toward a subfield of the l$\sim59\degr$ Hi-Gal SDP field, we performed a pixel
by pixel fitting of the full dust SED between 8 and 500\,$\mu$m using a 
physical dust emission model, {\tt DustEM} \citep{compiegne2010}. 

Assuming that the BG properties remain constant,
the unique wavelength coverage provided 
by the {\it Spitzer} and {\it Herschel} photometric observations allows us to 
derive the following parameters for our dust model : the PAH and VSG 
abundances, $Y_{PAH}$ and $Y_{VSG}$, the column density, $N_H$, and the 
intensity of the exciting radiation field, $U_{MMP83}$. 
To our knowledge, this is the first time PAH and VSG abundance maps are 
derived using such a pixel by pixel SED fitting at resolution 
$<1\arcmin$ and over such an extended field. These abundances, as well as 
$N_H$ and $U_{MMP83}$, vary significantly over the field. 
As already reported by previous studies, $Y_{PAH}$ and $Y_{VSG}$ appears 
not to be positively correlated.

Although it was already theoretically known, our analysis method provide a 
firm demonstration that IRAC\,$8\,\mu$m does not trace the PAH abundance but 
the product $Y_{PAH} \times N_H \times U_{MMP83}$. 
We also showed that at $70\,\mu$m, the modeled emission is due to both 
the BG and the VSG.  At these wavelengths the BG emission is sensitive to 
both the intensity of the exciting radiation field and the column density 
likewise the VSG, explaining the similar spatial structure seen in 
PACS\,$70\mu$m maps regarding shorter wavelengths. Using our model, we derived
the VSG contribution to the PACS channels that can be up to $\sim50\%$, 
$\sim 17\%$ and $\sim 7\%$ at 70, 100 and $\,160\,\mu$m, respectively. 
We conclude that the interpretation of {\it Herschel} spectrum of the diffuse 
emission down to 70\,$\mu$m may require to take into account the 
stochastically heated population (VSG).

Finally, our analysis allows for a better understanding of the first 
{\it Herschel} images of the galactic plan diffuse emission by disentangling 
between the different dust population contributions and also revealing the 
great potential of the {\it Herschel}/{\it Spitzer} synergy combined with a 
physical dust model for the study of dust evolution \citep[see also][]
{abergel2010}.

\bibliographystyle{apj}

\end{document}